\documentclass[twocolumn,showpacs,preprintnumbers,amsmath,A4paper]{revtex4}

\usepackage{graphicx}% Include figure files
\usepackage{dcolumn}% Align table columns on decimal point
\usepackage{bm}% bold math

\begin{document}
\newcommand{\kp}{{\bf k$\cdot$p}\ }
\newcommand{\Pp}{{\bf P$\cdot$p}\ }

%\preprint{APS/123-QED}
\title{Wave and uncertainty properties of electrons in crystalline solids}
\author{ Wlodek Zawadzki}
 \affiliation{Institute of Physics, Polish Academy of Sciences,\\
Al.Lotnikow 32/46, 02-668 Warsaw, Poland\footnotetext{$^*$ e-mail address: zawad@ifpan.edu.pl}\\}
\

%\date{\today}% It is always \today, today,
             %  but any date may be explicitly specified
\begin{abstract}
Relations between particle and wave properties for charge carriers in periodic potentials of crystalline metals and semiconductors are derived. The particle aspects of electrons and holes in periodic potentials are considered using properties of quasimomentum (QM), while the wave aspects are described employing wave packets of Bloch waves. The two aspects are combined in the derivation of QM-wavelength relations for energy bands of arbitrary nonparabolicity and nonsphericity. An effective mass relating electron QM to its average velocity for spherical energy bands is defined and used to calculate energy dependences of wavelengths for electrons in narrow gap semiconductors, graphene and surface states of topological insulators. An uncertainty relation between electron quasimomentum and its spatial coordinate in periodic potentials is derived. It is emphasized that the described properties apply to the average (not instantaneous) electron behavior. Analogies between the wave and uncertainty properties of electrons in crystalline solids and those in vacuum are traced.
\end{abstract}

\maketitle

\section{\label{sec:level1}INTRODUCTION\protect\\ \lowercase{}}

Our work is concerned with the wave properties of electrons moving in periodic potentials of crystal lattices. De Broglie [1] showed that a free quantum particle in vacuum has wave properties having the wavelength $\lambda=h/p$, where $p$ is particle's momentum. We want to derive a corresponding formula for an electron in a crystalline solid. In our understanding such a derivation does not exist. Still, one uses for solid-state electrons or holes a relation similar to that derived by de Broglie based on the following implicit reasoning. For an electron in a periodic potential, solutions of the Schrodinger equation are given by the Bloch functions $\Psi=\exp(i\textbf{k}\cdot \textbf{r})u_\textbf{k}(\textbf{r})$, where $u_\textbf{k}(\textbf{r})$ are amplitudes having the periodicity of the potential. The quantity $\textbf{k}$ is called the wave vector and it is usually stated that $\hbar \textbf{k}$ corresponds to electron's quasimomentum. On the other hand, since by its name the wave vector should be $k=2\pi/\lambda$, one quickly obtains the desired relation similar to that of de Broglie.

The above reasoning, however, raises several objections. First, the quantity $\textbf{k}$ in the Bloch function is \emph{a priori} only a label or quantum number to designate the eigenfunctions of translations, so its relation to electron's quasimomentum is not obvious. Second, it is the \emph{name of the wave vector} alone (sometimes called the wave number) which \emph{suggests} its relation to the wavelength $\lambda$. This relation is to be derived. Third, the periodic amplitude $u_\textbf{k}(\textbf{r})$ (resulting from the presence of periodic potential) affects the total $\textbf{r}$-dependence of the Bloch function, so the latter involves in general more than one value of $\textbf{k}$. Finally, an electron moving in a periodic potential experiences a Zitterbewegung (trembling motion) [2, 3, 4], while the above reasoning ignores this effect. One can rephrase the last point by observing that the instantaneous velocity, related to the standard momentum $\textbf{p}$, is not a constant of the motion. These objections have motivated us to a more rigorous consideration of the wave properties related to charge carriers in crystalline solids. This problem is closely related to the uncertainty principle and we treat the latter as well.

In order to describe the wave behavior of a particle one needs to consider separately its particle and wave aspects, and then to combine the two.
In his historic work, de Broglie used the invariance of a wave's phase in special relativity (see also ref. [5]).
Treating the problem for crystalline solids one cannot use this procedure because electron velocities in metals or semiconductors are nonrelativistic. On the other hand, such a derivation should be valid for an energy band of arbitrary energy-momentum dispersion. Bohm [6] proposed for particles in vacuum an approach alternative to that of de Broglie, not based on special relativity. A similar but somewhat more general reasoning was given by Messiah [7]. Considering the particle aspect, we treat in some detail the notion of quasimomentum (also called in the literature pseudo-momentum or crystal momentum). In order to avoid the often-used but misleading associations mentioned above, we do not use the letter $\textbf{k}$ and related names "wavevector" or "wavenumber". In order to keep the main derivation compact, we do not derive known auxiliary results, but write them down and quote appropriate references.

The paper is organized as follows. In section II A we describe particle aspects of electrons in periodic potentials. In section II B we follow with the wave aspects, combine the two and obtain final results. In section III we treat the uncertainty principle. Section IV contains the discussion, the paper is concluded by the summary. In Appendix we present a somewhat different version of the wave aspect and final reasoning.

\section{\label{sec:level2}WAVE PROPERTIES OF ELECTRONS IN PERIODIC POTENTIALS \lowercase{}}

In this section we derive an analog of the de Broglie relation for electrons in periodic potentials . Our considerations are divided into the particle aspect and the wave aspect of the problem.

\subsection{ Particle aspect}

The Schrodinger equation for an electron in a periodic potential $V(\textbf{r})$ of the lattice is
\begin{equation}
{\hat H}\Psi_{n}={\cal E}_{n}\Psi_{n}\;,
\end{equation}
with the Hamiltonian
\begin{equation}
{\hat H}=\frac{p^2}{2m_0}+V(\textbf{r})\;,
\end{equation}
where $n$ is the band index and $m_0$ is the free electron mass. The solutions of Eqs. (1) and (2 ) are given by the Bloch functions ($n$ is omitted)
\begin{equation}
\Psi_{\textbf{P}}(\textbf{r})=e^{i\textbf{P}\cdot \textbf{r}/\hbar}u_\textbf{P}(\textbf{r})\;.
\end{equation}

Because of the translation symmetry of the Hamiltonian there must exist a physical quantity that is conserved during the electron motion corresponding to this symmetry. One names this quantity quasimomentum (QM) and we denote it as $\textbf{P}$. Its meaning and properties are to be determined. Since QM is supposed to be a constant of the motion, its operator must commute with the Hamiltonian
\begin{equation}
\bm{\hat{P}}{\hat H}-{\hat H}\bm{\hat{ P}}=0\;,
\end{equation}
and the Bloch functions are also the eigenfunctions of $\bm{\hat{ P}}$
\begin{equation}
\bm{\hat{P}}\Psi_{\textbf{P}}(\textbf{r}) = \textbf{P}\Psi_{\textbf{P}}(\textbf{r})\;.
\end{equation}
We look for the operator $\bm{\hat{ P}}$ in the form [8, 9]
\begin{equation}
\bm{\hat{P}}=-i\hbar\bm{\nabla_r}+i\hbar \bm{\gamma}(\textbf{r})\;,
\end{equation}
where $\bm{\gamma}(\textbf{r})$ is a function having the periodicity of the lattice. Equation (5), accounting for Eq. (3), gives
$$
\bm{\hat{P}}\Psi_{\textbf{P}}=-i\hbar\frac{i}{\hbar}\textbf{P}\Psi_{\textbf{P}}+e^{i\textbf{P}\cdot \textbf{r}/\hbar}(-i\hbar\bm{\nabla}u_{\textbf{P}}+i\hbar \bm{\gamma}\Psi_{\textbf{P}})=
$$
\begin{equation}
=\bm{\hat{P}}\Psi_{\textbf{P}}+i\hbar(\bm{\gamma}-\bm{\nabla}\ln|u_{\textbf{P}}|)\Psi_{\textbf{P}}=\textbf{P}\Psi_{\textbf{P}}\;,
\end{equation}
provided that $\bm{\gamma}=\bm{\nabla} \ln |u_{\textbf{P}}|$. In consequence, the quasimomentum operator is (the band index $n$ is restored)
\begin{equation}
\bm{\hat{P}}_n(\textbf{r})=-i\hbar\bm{\nabla}+i\hbar[{\bm\nabla} \ln |u_{n\textbf{P}}(\textbf{r})|]\;.
\end{equation}
The square parenthesis indicates that $\nabla$ acts only on the subsequent expression. In contrast to the standard momentum, the QM operator $\bm{\hat{P}}_n$ is different for different bands. Since the energy and QM can be measured simultaneously, one can plot ${\cal{E}}_{n}(\textbf{P})$ which express the band dispersions.

Next we consider electron velocity in a band. An instantaneous electron velocity $\bm{\hat{v}}=\bm{\hat{p}}/m_0$ is not a constant of the motion in the presence of a periodic potential (or any electric potential). On the other hand, the average velocity in a band
\begin{equation}
\bm{\overline{v}}_n(\textbf{P})=<\Psi_{n\textbf{P}}|\frac{\bm{\hat{p}}}{m_0}|\Psi_{n\textbf{P}}>\;.
\end{equation}
is a constant of the motion. This average velocity can be expressed by the well known formula [9, 10, 11, 12, 13, 14] (n is suppressed)
\begin{equation}
\bm{\overline{v}}(\textbf{P})=\bm{\nabla}_{\textbf{P}}\cal{E}\;.
\end{equation}

The above formula, which is valid for bands of any dispersions and nonsphericities, is all we need to proceed with the further derivation. However, we develop somewhat further the particle aspect for a more restricted case of electrons in spherical energy bands, as it is useful in applying final relations to specific examples and in comparisons with vacuum.

We \emph{define an electron effective mass} relating the average velocity to quasimomentum
\begin{equation}
\hat{m^*}\bm{\overline{v}}=\textbf{P}\;.
\end{equation}
where $\hat{m^*}$ is in principle a 3 x 3 tensor. We emphasize that, for a band of nonquadratic dispersion, the above "velocity mass" is not the same as the "acceleration mass" relating acceleration to force, as commonly introduced in textbooks. By using Eq. (10) one calculates for a spherical band
\begin{equation}
\overline{v}_i=\frac{\partial \cal{E}}{\partial P_i}=\frac{d\cal{E}}{dP}\frac{\partial P}{\partial P_i}=\frac{d\cal{E}}{dP}\frac{P_i}{P}=
\frac{d\cal{E}}{dP}\frac{1}{P}\delta_{ij}P_j\;,
\end{equation}
where the last term uses the sum convention for the repeated coordinate subscript. Using the definition (11), the inverse mass tensor is
\begin{equation}
\overline{v}_i=\left(\frac{1}{m^*}\right)_{ij}P_j\;.
\end{equation}
By employing Eq. (12) one obtains
\begin{equation}
\left(\frac{1}{m^*}\right)_{ij}=
\frac{d\cal{E}}{dP}\frac{1}{P}\delta_{ij}\;,
\end{equation}
which shows that the inverse mass tensor is a scalar for a spherical band
\begin{equation}
\frac{1}{m^*}=\frac{1}{P}\frac{d\cal{E}}{dP}\;.
\end{equation}
Consequently, the average velocity is, see Eq. (11),
\begin{equation}
\bm{\overline{v}}=\frac{\textbf{P}}{m^*}\;,
\end{equation}
which is valid also for the absolute values
\begin{equation}
{\overline{v}}=\frac{{P}}{m^*}\;.
\end{equation}

Finally, combining Eqs. (9) and (16) we obtain
\begin{equation}
\frac{\overline{{\textbf{p}}}}{m_0}=\frac{{\textbf{P}}}{m^*}\;,
\end{equation}
where ${\overline{\textbf{p}}}$ is the average value of canonical momentum in a given Bloch state and $\textbf{P}$ is the value of quasimomentum in the same state. Since usually there is $m^* << m_0$, it follows that ${\overline{\textbf{p}}}$ is much larger than $\textbf{P}$.

\subsection{ Wave aspect. de Broglie-type relations}

To complete the derivation one needs to consider the wave aspect of electrons in a periodic potential. The important step is to show that the average electron velocity in a crystal, as considered above, is equal to the group velocity of a wave packet constructed from the Bloch waves. The question of this equality is controversial [12], [15], so we consider it in some detail. On the other hand, this point is essential since it introduces the wavelength appearing in the de Broglie-type relations. Also, the construction of the packet is necessary because a single Bloch wave does not represent a point-like moving electron, while maximum of a packet can be associated with a moving particle.

 We present two versions of the wave aspect and derivation of final results. The two versions are similar but not identical. We give preference to the one given directly below because it is more adapted to crystalline solids. The other version, closer to those of Bohm [6] and Messiah [7] for vacuum, is given in the Appendix. Final results of the two versions are the same.

In order to avoid writing the Planck constant too often we introduce the notation $\textbf{P}/\hbar = \bm{\beta}$. A packet of Bloch monochromatic waves belonging to one band is given by the integral

\begin{equation}
\Phi(\textbf{r}, t)=\int^{\bm{\beta_0}+\Delta\bm{\beta}}_{\bm{\beta_0}-\Delta\bm{\beta}} f({\bm{\beta}}-\bm{\beta_0})e^{i(\bm{\beta}\cdot\textbf{r}-\omega t)}u_{\bm{\beta}}(\textbf{r})d\beta_x d\beta_y d\beta_z\;,
\end{equation}
where $f(\bm{\beta})$ is the weight of Bloch waves contributing to the packet. The value $\bm{\beta_0}$ determines the central wavelength of the group and $\Delta \bm{\beta} = \bm{\beta} - \bm{\beta_0}$ is supposed to be small. The frequency $\omega$ depends in general on $\bm{\beta}$. We want to take the periodic amplitude $u_{\bm{\beta}}$ out of the integral sign and the question arises when it may be done. It is clear that the desired procedure gains on validity when $\Delta \bm{\beta}$ is small since then $u_{\bm{\beta}}$ changes only slightly.

To get a more precise evaluation we recall that, because of the periodicity of $u_{\bm{\beta}}(\textbf{r})$, the latter can be developed into the Fourier series of plain waves

\begin{equation}
u_{\bm{\beta}}(\textbf{r})=\sum_{\nu} c_{\nu}({\bm{\beta}})\exp(i\textbf{b}_{\nu}\cdot \textbf{r})\;,
\end{equation}
where $c_{\nu}({\bm{\beta}})$ are coefficients and $\textbf{b}_{\nu}$ are reciprocal lattice vectors, see [10, 11, 12, 16, 17, 18]. The index $\nu$ takes negative and positive integer values including zero. For weak periodic potentials the coefficients $c_{\nu}$ can be calculated explicitly and it turns out that they decrease as functions of growing $\nu$ [11, 12, 16, 17, 18, 19]. For this case there is $c_{0}\approx 1$ since, as the potential goes to zero, the Bloch wave should reduce to the plain wave. Let us write the Bloch wave using Eq. (20) and including only $\nu = -1, 0, +1$.

\begin{equation}
e^{i\bm{\beta}\cdot\textbf{r}} u_{\bm{\beta}}(\textbf{r})\approx
e^{i\bm{\beta}\cdot\textbf{r}}
(c_{-1}e^{i\textbf{b}_{-1}\cdot\textbf{r}}+c_0+
c_1e^{{i\textbf{b}_1} \cdot \textbf{r}})\;.
\end{equation}
One can now see explicitly that, for the above approximation, one has three effective $\bm{\beta}$ values involved in the packet. The additional two values come from the presence of the periodic Bloch amplitude. Considering for simplicity only one direction $x$ and taking $b_{\pm1}=\pm2\pi/a$, where $a$ is the lattice period, one has on both sides of the central value ${\beta}_{0}$ two additional values $\beta_{0} \pm2\pi/a$.
Going back to the integral (19) it is clear that, if the weight function is symmetric around $\beta_{0}$ and $\Delta \beta$ is narrower than $2\pi/a$, the side terms in Eq. (21) are excluded and only the middle term with $c_0$ remains. The above procedure can be easily generalized to three dimensions. Thus, for weak periodic potentials, $u_{\bm{\beta}}(\textbf{r})$ can be legitimately taken out of the integral sign. For strong periodic potentials, however, while the Fourier expansion of Eq. (20) is still valid, the coefficient $c_0$ will in general depend on $\bm{\beta}$, so taking $u_{\bm{\beta}}$ out of the integral becomes approximate.

With this reservation in mind, once $u_{\bm{\beta}}$ is taken out of the integral sign, the resulting wave packet is
\begin{equation}
\Phi(\textbf{r}, t)=c_0\int^{\bm{\beta_0}+\Delta\bm{\beta}}_{\bm{\beta_0}-\Delta\bm{\beta}} f({\bm{\beta}}-\bm{\beta_0})e^{i(\bm{\beta}\cdot\textbf{r}-\omega t)}d\beta_xd\beta_yd\beta_z\;.
\end{equation}
This is a standard wave packet of monochromatic plain waves, so the group velocity of its maximum is given by the well-known formula
\begin{equation}
\bm{v}_{g}=\bm{\nabla}_{\bm{\beta}} \omega=\bm{\nabla}_{\textbf{P}}\cal{E}\;,
\end{equation}
if we write ${\cal{E}}=\hbar \omega$. Comparing Eq. (23) with Eq. (10) it is seen that, in the above approximation, the group velocity of the packet is equal to the average electron velocity, i.e.
\begin{equation}
\overline{\bm{v}}={\bm{\nabla}_{\textbf{P}}}{\cal{E}}=\bm{v}_{g}\;,
\end{equation}
where the gradient should be taken at $\textbf{P} = \textbf{P}_0$. On the other hand, in the plane waves appearing in Eq. (22) there is $\bm{\beta}_i = 2\pi/\lambda_i$, where $\lambda_i$ are wavelengths. Consequently
\begin{equation}
P_i= \frac{h}{\lambda_i}\;\;\;{\rm{or}}\;\;\;\lambda_i=\frac{h}{P_i}\;.
\end{equation}

The above equations are our final results. They resemble the classic de Broglie relations between particle and wave characteristics for vacuum, the important difference being that the vector of canonical momentum $\textbf{p}$ for vacuum is replaced by the vector of electron quasimomentum $\textbf{P}$ for periodic potentials. The results (25) have been obtained for arbitrary energy bands, both nonparabolic and nonspherical. The wavelengths $\lambda_i$ are those of maxima of the wave packet associated with the electron.

It is seen that the relations (25) have vector forms, similarly to vacuum. For vacuum one usually writes $\lambda=h/p$ with the use of absolute values. This can be done because in vacuum one can choose the coordinate system in such a way that only one component of the momentum is nonzero. In a crystal this cannot be done so easily since a three-dimensional periodic potential of the lattice is fixed in space , so that a moving electron has in general three different QM components $P_i$. However, according to the first Eq. (25), vectors $P_i$ and $h/\lambda_i$ have equal components, so their absolute values and directions are the same. This gives for the absolute values

\begin{equation}
P=\frac{h}{\lambda}\;\; {\rm{or}} \;\;\lambda=\frac{h}{P}\;,
\end{equation}
where
\begin{equation}
\frac{1}{\lambda}=\left[\left(\frac{1}{\lambda_x}\right)^2+\left(\frac{1}{\lambda_y}\right)^2+\left(\frac{1}{\lambda_z}\right)^2\right]^{1/2}\;,
\end{equation}
is the "effective" wavelength given by the three components. Thus, we obtain a similar relation between $\lambda$ and $P$ as for vacuum between $\lambda$ and $p$. If the motion is in one direction only, one deals with the simple absolute values $\lambda$ and $P$.

In conclusion it should be mentioned that, in the above procedure, we have assumed the quantum relation between the frequency and energy. We come back to this point in the discussion.

\section{\label{sec:level2} APPLICATIONS \lowercase{}}

\begin{figure}
\includegraphics[scale=0.36,angle=0, bb = 500 45 222 560]{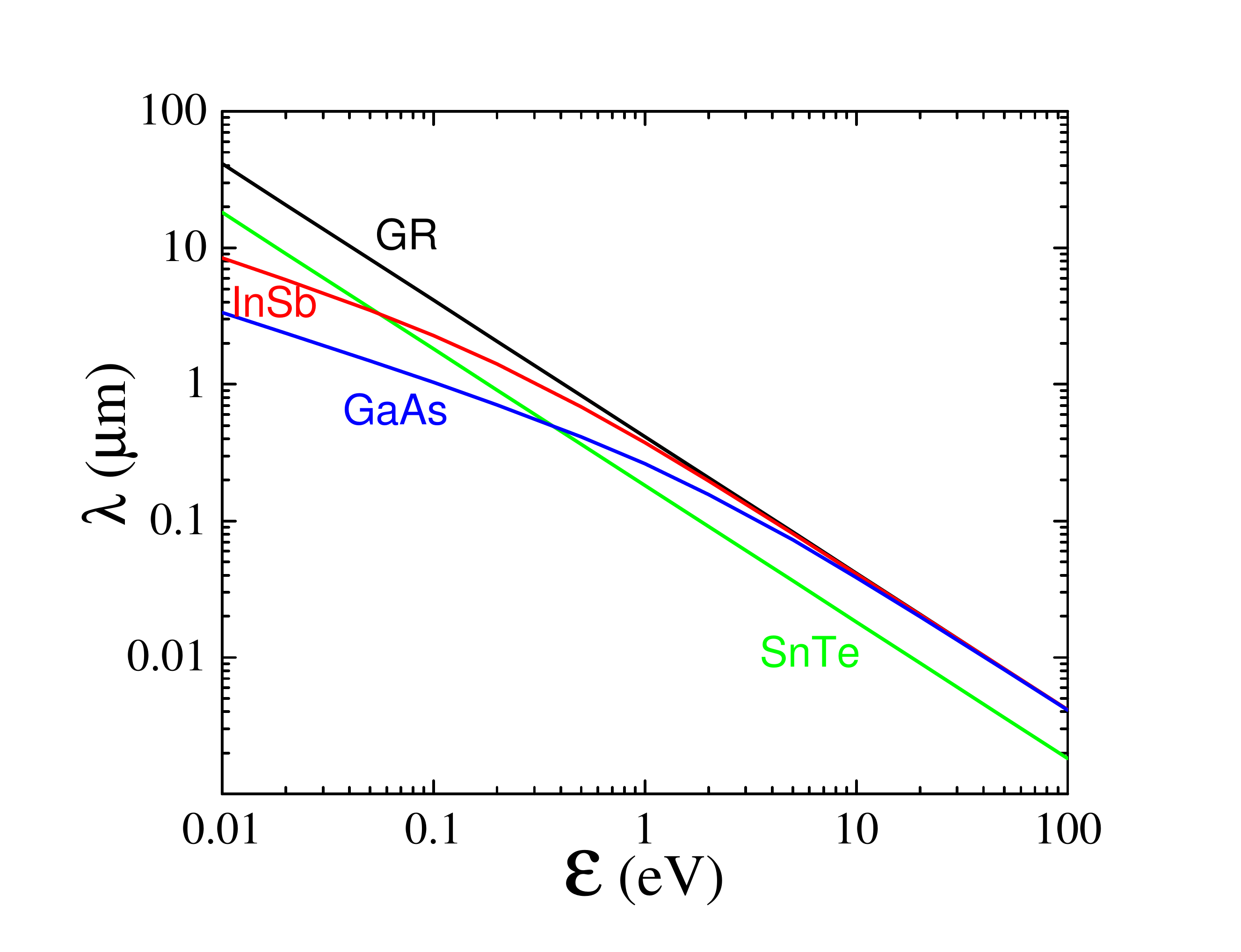}
\caption{\label{fig:epsart}{ Wavelength $\lambda$ versus electron energy ${\cal{E}}$ calculated using the two-band model of Eq. (32) for four materials. The velocity $u = 1\times 10^8$ cm/s for InSb (${\cal{E}}_g$ = 0.23 eV), GaAs (${\cal{E}}_g$ = 1.5 eV), GR (${\cal{E}}_g$ = 0), and $u = 4.4\times 10^7$ cm/s for topological surface states of SnTe (${\cal{E}}_g$ =0), see Ref. [23].}} \label{fig1th}
\end{figure}

Here we apply the derived relations to standard and less standard energy bands. For the standard parabolic and spherical band there is ${\cal{E}}=P^2/2m^*_0$, where $\cal{E}$ is the electron energy counted from the band edge and $m^*_0$ is the effective mass at the band edge. In consequence
\begin{equation}
\lambda=\frac{h}{(2m^*_0 {\cal{E}})^{1/2}}\;.
\end{equation}
In the presence of a slowly varying external potential $U(\textbf{r})$ there is
\begin{equation}
\lambda=\frac{h}{[2m^*_0({\cal{E}}-U)]^{1/2}}\;.
\end{equation}
The above expressions have their correspondence for electrons in vacuum with the effective mass replaced by the free electron mass.

For a spherical band of arbitrary dispersion ${\cal{E}}(P)$ one has, see Eq. (16),
\begin{equation}
\lambda=\frac{h}{m^*{\overline v}}\;,
\end{equation}
where $\overline v$ is the average electron velocity and $m^*(\cal{E})$ is the energy-dependent velocity mass. Formula (30) is in analogy to that for vacuum, see discussion.

Next we consider more specifically spherical but nonparabolic energy bands of narrow-gap semiconductors. In narrow-gap III-V and II-VI compounds the relation $\cal{E}(P)$ is described by the so called two-band model (2BM), see [20, 21],
\begin{equation}
{\cal{E}}({\cal{E}}_g+{\cal{E}})=u^2P^2\;,
\end{equation}
where ${\cal{E}}_g$ is the energy gap and $u=({\cal{E}}_g/2m^*_0)^{1/2}=10^8$ cm/s is the characteristic velocity having very similar value for various materials. For energies near the band edge, the bands described by 2BM are parabolic, while for higher energies they become linear [21]. It follows from Eq. (31)
\begin{equation}
\lambda=\frac{hu}{[{\cal{E}}({\cal{E}}_g+{\cal{E}})]^{1/2}}\;.
\end{equation}
For low energies ${\cal{E}}<<{\cal{E}}_g$ one obtains
\begin{equation}
\lambda=\frac{hu}{({\cal{E}}{\cal{E}}_g)^{1/2}}\;,
\end{equation}
which is equivalent to Eq. (28). On the other hand, for high energies ${\cal{E}}>>{\cal{E}}_g$ we get
\begin{equation}
\lambda=\frac{hu}{\cal{E}}\;,
\end{equation}

The two-band model of Eq. (31) describes correctly also the zero-gap situation ${\cal{E}}_g$=0. One then obtains ${\cal{E}}=uP$. This is the case of Hg$_{0.835}$Cd$_{0.165}$Te in three dimensions and of graphene in two dimensions [21, 22]. In these two cases formula (34) applies. Remarkably, also for graphene the velocity $u$ has almost exactly the value given above, although the symmetry of this material is different from those of III-V and II-VI compounds. In addition, one can describe linear dispersions of the surface states in topological insulators by ${\cal{E}}=uP$ and use Eq. (34), but the velocity $u$ (often called the Fermi velocity) should be adjusted for each material.

\section{\label{sec:level2} UNCERTAINTY PRINCIPLE\protect\\ \lowercase{}}

The uncertainty principle for particle momentum and its spatial coordinate in vacuum is usually derived using wave packets or, more rigorously, noncommutativity properties of the corresponding operators, see e.g. [24, 25]. We give the wave aspect of the corresponding derivation for crystalline solids and then quote the rigorous result.

We begin with Eq. (22) describing an electron by the wave packet of monochromatic Bloch waves in which the periodic amplitude is taken approximately out of the integral sign and replaced by a constant. The electron motion is assumed to be along the $x$ direction. As above, we use the notation $\beta = P/\hbar$ and write $\beta = \beta_0+ \Delta\beta$ where $\Delta\beta = \beta - \beta_0$ is assumed to be small. In consequence, the frequency $\omega$ can be developed
\begin{equation}
\omega = \omega_0+ \frac{d\omega}{d\beta}(\beta-\beta_0) = \omega_0 + v_g\Delta\beta\;,
\end{equation}
where $v_g$ is the group velocity and the frequency $\omega_0$ corresponds to $\beta_0$. The weight $f(\beta)$ is usually a slowly varying function of $\beta$ (in the narrow range of $\Delta\beta$ it can even be a constant). In this case one can assume $f(\beta)\approx f(\beta_0)$ and take it out of the integral sign. Introducing a new variable of integration $\xi = \Delta \beta$ and incorporating the above relations one obtains the integral in the form
\begin{equation}
\Phi(x, t)=c_0f(\beta_0)e^{i(\beta_0 x-\omega_0 t)}\int^{+\Delta{\beta}}_{-\Delta\bm{\beta}} e^{i(x-v_g t)\xi}d\xi\;.
\end{equation}
This can be integrated to give
$$
\Phi(x, t)=2c_0f(\beta_0)\frac{sin[(x-v_g t)\Delta \beta]}{x-v_g t}e^{i(\beta_0 x-\omega_0 t)}=
$$
\begin{equation}
=g(x, t)e^{i(\beta_0 x-\omega_0 t)}\;.
\end{equation}
Because $\Delta{\beta}$ is supposed to be small, $g(x, t)$ changes slowly as a function of $x$ and $t$, so that it can be considered as an amplitude of a monochromatic wave and $\beta_0 x-\omega_0 t$ as its phase. Let us consider $t=0$. For $x = 0$ the amplitude is at its maximum. A spatial extension $\Delta x$ of the wave packet $\Phi(x, 0)$ can be taken to be a double distance $x_1$ between the maximum and its first zero at $x_1 \Delta \beta = \pi$. This gives the estimation $\Delta \beta \Delta x= 2\pi$ and finally
\begin{equation}
\Delta P_x \Delta x \approx 2\pi\hbar\;.
\end{equation}

A rigorous result for the uncertainty relation can be obtained by considering the noncommutativity of $\hat{x}$ and $\hat{P_x}$ operators. The commutator $[\hat{x}, \hat{P_x}] = i\hbar$ has the same value as the commutator of $\hat{x}$ and $\hat{p_x}$, see Eq. (8). In consequence, the uncertainty inequality is also the same. Thus we have
\begin{equation}
\Delta P_x \Delta x \ge \hbar/2\;.
\end{equation}
This result is analogous to the well-known Heisenberg inequality for particles in vacuum with the quasimomentum ${P_x}$
replacing the canonical momentum ${p_x}$.

\section{\label{sec:level2} DISCUSSION\protect\\ \lowercase{}}

As mentioned in the introduction, the de Broglie-type relations between particle and wave characteristics have been taken for granted without much derivation for charge carriers in periodic potentials of crystal lattices. The relations derived above confirm the common belief. We also enumerated objections to the simple reasoning leading to this belief. Our considerations deal explicitly or implicitly with these objections. And so, in the particle aspect it is shown that the quantum number appearing in the Bloch function is the quasimomentum $\textbf{P}$ representing a constant of the motion related to the periodicity of the potential. This point is related to the phenomenon of the trembling motion (Zitterbewegung, ZB) present during the electron propagation through the periodic potential, see [2, 4]. The trembling is a consequence of the fact that the canonical momentum $\textbf{p}$ is not a constant of the motion in the presence of a periodic potential. Thus the derived de Broglie-type relation in solids, involving quasimomentum $\textbf{P}$, is constant in time and \emph{ valid on average}. Simultaneously, there exists another (standard) de Broglie relation, related to momentum $\textbf{p}$, being valid instantaneously. In the presence of ZB, when the electron velocity trembles in time, so trembles the corresponding wavelength.

We further showed under what restrictions one may neglect other wavelengths related to the presence of the periodic Bloch amplitude. This question remains somewhat ambiguous because one cannot control the width of a wave packet representing the solid-state electron. Generally speaking, the particle aspect of electron can be described rigorously, while the wave aspect, as described by a packet, has an approximate character. The way to deal with a packet of Bloch waves is to reduce it to a packet of plain waves. This can be done to a good approximation for weak periodic potentials, while for stronger potentials the required approximation becomes progressively worse.

An important step in the derivation is to assume the quantum relation between the frequency of the monochromatic Bloch waves and the electron energy $\hbar \omega=\cal{E}$. In his original derivation, de Broglie used special relativity and required an invariance of the wave's phase in two coordinate systems to obtain both relations: $\hbar \omega=\cal{E}$ and $\lambda=h/p$, see Wichmann [5]. The problem with electrons in solids is that one cannot use relativity because, on the one hand, electron velocities are nonrelativistic and, on the other, one looks for a derivation valid for arbitrary band dispersion and nonsphericity, i.e. for situations considerably more general than occur in vacuum. However, the quantum Ansatz used above should not be considered doubtful since in solids one deals in the vast majority of cases with the domain of quantum physics.

The derived wave properties of electrons in solids resemble the corresponding de Broglie relations for vacuum, if the canonical momentum $\textbf{p}$ is replaced by the quasimomentum $\textbf{P}$. It is well known that $\textbf{P}$ often (but not always!) plays for solids the role that $\textbf{p}$ plays for vacuum. One can push this analogy further by observing that the two-band model for narrow gap semiconductors, introduced in Eq. (31), is in close analogy to special relativity with the correspondence: ${\cal{E}}_g \rightarrow 2m_0c^2$ and $m^*_0 \rightarrow m_0$, see [21]. It is then not surprising that using this model to describe applications of the de Broglie-type relations to NGS, one obtains formulas in close analogy to vacuum. And so Eqs. (32), (33), (34) are identical with those for vacuum, if the above correspondence is used, see [26]. Formula (30), relating the wavelength to the energy-dependent effective mass and average velocity, deserves a comment since the analogous relativistic formula for vacuum was a subject of some controversy. Namely, it was disputed whether the mass entering this formula should be the rest particle mass or the velocity (or energy) dependent mass. It was indicated that the second formulation was correct, see [27]. This is again similar to Eq. (30) which contains the energy dependent effective electron mass.

Concerning the uncertainly principle, it was explicitly demonstrated for a superlattice of finite length that the resulting uncertainty is $\Delta x \Delta P_x \approx 4\pi \hbar$, see [28]. We emphasize that, in the derivation of the uncertainty principle for solids it was useful to employ a packet of Bloch waves in order to show that it is again the quasimomentum $\textbf{P}$ that is involved in the analog of the well known Heisenberg principle for vacuum. This procedure is subject to the approximations mentioned above. Fortunately, there exists the exact and well known derivation of the corresponding inequality (not quoted above) with the result given in Eq. (39). However, the exact calculation does not allow one to decide \emph{a priori} whether it is $p_x$ or $P_x$ that enters the inequality for solids, since the noncommutativity of $\hat{x}$ and $\hat{P_x}$ is the same as that of $\hat{x}$ and $\hat{p_x}$.

\section{\label{sec:level2} SUMMARY\protect\\ \lowercase{}}

De Broglie-type relations between particle and wave characteristics of electrons in periodic potentials of crystalline solids are derived. We clarify objections arising when one tries to use automatically intuitive relations existing in the literature. Approximations involved in the derivation are noted. The obtained relations are valid for energy bands of arbitrary nonsphericity and nonparabolicity in the quasimomentum space and they are applied to spherical bands of narrow gap semiconductors, graphene and surface states of topological insulators. The uncertainty principle is also considered for electrons in periodic potentials. One can summarize our practical results with the statement that the relations for periodic potentials correspond to those in vacuum in which the canonical momentum $\textbf{p}$ is replaced by the quasimomentum $\textbf{P}$.

\begin{acknowledgments}

I am obliged to Dr Pawel Pfeffer for elucidating discussions.

\end{acknowledgments}
\appendix*
\section{}

Here we briefly present an alternative version of the wave aspect necessary for the derivation of the de Broglie-type relations. As compared to the version given above in section II B, the main difference is that we describe the Bloch waves introducing a new vector $\textbf{q}$, whose relation to the quasimomentum is in principle unknown. (The vector $\textbf{q}$ should not be confused with the vector $\bm{\beta}$ introduced above, which only served to simplify the notation.) This alternative procedure requires a modification of the final reasoning. The version below is closer to the nonrelativistic derivations for vacuum, see [6, 7]. Final results of the two versions are the same.

The packet of monochromatic Bloch waves is now, c.f. Eq. (19)

\begin{equation}
\Phi(\textbf{r}, t)=\int^{\textbf{q}_0+\Delta\textbf{q}}_{\textbf{q}_0-\Delta\textbf{q}} f(\textbf{q}-\textbf{q}_0)e^{i({\textbf{q}}\cdot\textbf{r}-\omega t)}u_{\textbf{q}}(\textbf{r})dq_xdq_ydq_z\;,
\end{equation}
where, as previously, $f(\textbf{q})$ is the weight of waves with different $\textbf{q}$ contributing to the packet and $\omega$
depends in general on $\textbf{q}$. The periodic Bloch amplitude can again be developed into the Fourier series of plain waves, see Eq. (20),
\begin{equation}
u_{\textbf{q}}(\textbf{r})=\sum_{\nu} c_{\nu}(\textbf{q})\exp(i\textbf{b}_{\nu}\cdot \textbf{r})\;.
\end{equation}

Putting the above expression into Eq. (A1) one can reason as before that, for weak periodic potentials and the weight function symmetric around $q_0$
and $\Delta q$ smaller than $2\pi/a$ (for one dimension) , one can take the amplitude $u_{\textbf{q}}$ out of the integral sign. This transforms the integral (A1) into a packet of plain waves
\begin{equation}
\Phi(\textbf{r}, t)=c_0\int^{\textbf{q}_0+\Delta\textbf{q}}_{\textbf{q}_0-\Delta\textbf{q}} f(\textbf{q}-\textbf{q}_0)e^{i({\textbf{q}}\cdot\textbf{r}-\omega t)}dq_xdq_ydq_z\;.
\end{equation}

Writing $\hbar\omega = \cal{E}$, the group velocity of the packet is given by the formula
\begin{equation}
\bm{v}_{g}=\bm{\nabla}_{\textbf{q}} \omega=\bm{\nabla}_{\hbar\textbf{q}}\cal{E}\;.
\end{equation}
Comparing this with Eq. (10) one has $\overline{\bm{v}} = \bm{v}_g$ and
\begin{equation}
\textbf{P}=\hbar \textbf{q}\;\;,
\end{equation}
to within a constant vector, which one chooses equal to zero because of symmetry reasons, see [7].
On the other hand, for the plane waves in Eq. (A3) there is
\begin{equation}
q_i= \frac{2\pi}{\lambda_i}\;,
\end{equation}
so that
\begin{equation}
P_i=\frac{h}{\lambda_i}\;\;{\rm{or}}\;\;\lambda_i = \frac{h}{P_i}\;,
\end{equation}which is the final result identical with Eq. (25). As before, for strong periodic potentials the coefficient $c_0$ in Eq. (A.3) depends in general on $\textbf{q}$, so taking it out of the integral becomes approximate.

\end{document}